\documentclass[draft,12pt]{article}
\usepackage{amsmath,amsfonts,palatino,amsthm,epsf}
\newcommand{\beq}{\begin{equation}}
\newcommand{\eeq}{\end{equation}}

\newcommand{\f}{\begin{equation}}
\newcommand{\ff}{\end{equation}}
\newcommand{\blankline}{\vskip .3cm}

\begin{document}

\title{A quantization of topological $\cal M$ theory}
\author{   Lee Smolin\thanks{Email address:
lsmolin@perimeterinstitute.ca}\\
\\
\\
Perimeter Institute for Theoretical Physics,\\
35 King Street North, Waterloo, Ontario N2J 2W9, Canada, and \\
Department of Physics, University of Waterloo,\\
Waterloo, Ontario N2L 3G1, Canada\\}
\date{March 17, 2005}
\maketitle
\vfill
\begin{abstract}
A conjecture is made as to how to quantize topological $\cal M$ theory.
We study a Hamiltonian decomposition of Hitchin's $7$-dimensional
action and propose a formulation for it in terms of $13$ first class
constraints.  The theory has $2$ degrees of freedom per point,
and hence is diffeomorphism invariant, but not strictly speaking topological.
The result is argued to be equivalent to Hitchin's formulation.
The theory is quantized using loop quantum gravity methods.
An orthonormal basis for the diffeomorphism invariant states is given by
diffeomorphism classes of 
networks of two dimensional surfaces in the six dimensional manifold.
The hamiltonian constraint is polynomial and can be regulated by 
methods similar to those used in LQG.  

To connect topological $\cal M$ theory to full $\cal M$ theory,  a reduction from $11$ dimensional supergravity to Hitchin's $7$ dimensional theory is proposed.  One important conclusion is that the complex and symplectic structures represent non-commuting degrees of freedom.  This may have implications for attempts to construct phenomenologies on Calabi-Yau compactifications.

\end{abstract}
\vfill
\newpage
\tableofcontents

\section{Introduction}

Approaches to quantum gravity have so far fallen into two broad classes, according to whether they
are background independent or background dependent.  So far most work
on string and $\cal M$ theory has been based background dependent
methods and ideas.  But it has long been acknowledged that this was a temporary
expedient and that the ultimate principles of string theory must be formulated
in background independent terms.  Meanwhile, a great deal of progress has been made on background independent approaches, including loop quantum
gravity\cite{carlobook,invitation} , causal sets\cite{causalsets} and 
lorentzian dynamical triangulations\cite{LDT}.   

The results of these, especially loop quantum gravity (LQG), have inspired
a few attempts to approach string or $\cal M$ theory from a background
independent perspective\cite{d=11,bim}.  These make use of one of the 
most powerful observations of LQG, which is that theories of
gravity are closely related to topological field theories\cite{invitation}.
The precise relation is that gravitational theories are {\it constrained
topological field theories.} This means that their action is a sum
of the action for a $BF$ theory, plus quadratic constraints.  These are
sometimes called theories of forms, because the metric information is
coded into the dynamics of forms\cite{CDJ,twoform}\footnote{In $4$ spacetime dimensions, these 
turn out to be the self-dual two forms of a metric.}.
This is true  of general relativity in all dimensions\cite{higher}, 
as well as of supergravity in $11$ dimensions\cite{d=11}, so it is a fact
that must be relevant for how we formulate $\cal M$ theory.  

Recently Dijkgraaf et al\cite{DGNV} proposed a form of {\it topological
$\cal M$ theory}, which is a seven dimensional theory which is hypothesized to 
unify two six dimensional theories called 
{\it topological string theories\footnote{Related papers are \cite{related}.}.}  
This theory is defined by an action proposed by Hitchin\cite{hitchin,hitchin2}, 
and involves only
the dynamics of a three-form in seven dimensions.  Dijkgraaf et. al. in fact
propose that this theory is related by dimensional reduction to topological field theories relevant for three and four dimensional theories. This makes it natural
to suggest that the quantization of Hitchin's theory may be accomplished by
using background independent methods which have been successfully
applied to topological theories and theories of forms in lower dimensions.  

In this paper we make a first attempt at such a background independent quantization of topological $\cal M$ theory. In the next section we propose a form of the
theory as a constrained Hamiltonian theory.  
 We find that the dynamical variables are coded into a two form
$\beta$ and densitized bivector, $\pi$, on a compact six manifold $\Sigma$. 
These  are canonically conjugate to each other and are associated with the specification of two structures that go into 
the definition of a Calabi-Yau manifold, which are, respectively, a complex
and symplectic structure.  We find a system of first class constraints relating
them, which we argue is equivalent to the 
dynamics  described earlier by Hitchin in \cite{hitchin,hitchin2}.

In section 3 we count the local degrees of freedom, using 
standard methods.  We find there are two local degrees of freedom per point.
Thus, if the proposal made in this paper is correct, 
topological $\cal M$ theory is not actually a topological field theory. 

In section 4 we then quantize the local degrees of freedom, following the
methods of LQG.  We find a theory of extended objects living in the
six dimensional manifold, $\Sigma$. These are described by 
observables parameterized by membranes and  four 
dimensional branes in $\Sigma$. These involve, respectively, the complex
structure and symplectic structures on $\Sigma$.  
We find that the quantum states of the theory have a separable basis in 
one-to-one correspondence with the diffeomorphism equivalence classes of 
the membranes embedded in the six manifold.     

In the classical theory of Hitchin, the complex and symplectic  structures
each give a volume to $\Sigma$, and these are required to be equal
to each other.   In the Hamiltonian
formulation presented here, this condition is expressed by a hamiltonian 
constraint.  Its quantization leads to analogues of the  
Wheeler-deWitt equations. This has a form
not seen before, being cubic rather than quadratic in momenta. We are able
to use LQG methods to express the WdW operator as a limit of a sequence
of regulated operators. Unlike LQG, the operator is the sum of two
terms, and no easy solutions are apparent.  

Finally, in section 5, we show how the degrees of freedom of Hitchin's theory arise
from a dimensional reduction of $11$ dimensional supergravity in which 
the frame fields are set to zero.  

While these results may be seen as a first sketch of a quantum theory, there is
one intriguing question, raised in \cite{DGNV},  that  confronts us. The definition of a 
Calabi-Yau manifold requires fixing both the complex and symplectic structures.
Here we find that those structures do not commute with each other
quantum mechanically.  Thus, the use of Calabi-Yau manifolds to
describe compactifications of string and $\cal M$ theory can only be
sensible at a semiclassical level in which one works on a fixed,  
classical background
geometry.  Once quantum gravity effects are turned on, an uncertainty 
principle may prevent a quantum state as being identified as a Calabi-Yau
manifold.  This is true in Hitchin's theory, as pointed out in
\cite{DGNV}, but the fact that the degrees of 
freedom of that theory arise from a compactification of $11$ dimensional
supergravity suggest it will be true also in $\cal M$ theory.  

This gives rise to several fascinating questions that future work  may address. 

\begin{itemize}

\item{} Might there be quantum effects of order $l_{Pl}$ that arise
from the quantum fluctuations of the Calabi-Yau geometry? Could this
lead to new kinds of effects, perhaps observable in experiments
such as AUGER and GLAST?  

\item{}Might the quantum fluctuations in the Calabi-Yau geometries
help to stabilize then quantum mechanically against decay to the
negative energy density states found by \cite{unstable}?

\item{}If the Calabi-Yau compactifications do not correspond to quantum states
of the fundamental theory, but only arise in the classical limit,  there are implications for how they are to be counted in considerations of the landscape
of theories.  

\end{itemize}

\section{Hamiltonian formulation of Hitchin's theory}

Hitchin described a seven dimensional 
theory\cite{hitchin,hitchin2}, which Dijkgraaf et al propose
is a formulation of topological $\cal M$ theory\cite{DGNV}.  We begin
by reviewing their proposal. 

\subsection{Review of Topological $\cal M$ theory}

The theory is defined on a 
 $7$ dimensional manifold, $\cal M$, There is only one field, which is  
a real three form $\Omega$, with fixed cohomology class\footnote{\cite{DGNV} study
the theory in a dual formulation written in terms of a four-form on $\cal M$.}. 

Analogously to how the metric in LQG in 4d is formed from a set of two forms,
we can construct a metric on $\cal M$, $h(\Omega )$ depending only on
$\Omega$. As in the $4d$ case\cite{CDJ,twoform}, the densitized metric is cubic
in the form field. We have
\f
\tilde{h}_{ab}= \sqrt{h}h_{ab} = 
\Omega_{acd} \Omega_{bef} \Omega_{ghi} \epsilon^{cdefghi}
\ff
The action of Hitchin is
\f
I^{N}= \int_{\cal M}\sqrt{h(\Omega )}
\ff
This gives rise to a non-trivial theory when the cohomology class of $\Omega$ is
frozen. One then has
\f
\Omega = \Omega^0 + d\beta
\ff
for $\beta$ a two form.  With $\Omega^0$ fixed the  action is  a functional of 
$d \beta$ and hence generates an interesting dynamics.
\f
I^{N}[\beta ] = \int_{\cal M}\sqrt{h(\Omega^0 + d\beta )}
\ff

Put in this form, the action is invariant under gauge transformations
parameterized locally by a one form $\lambda$
\f
\beta \rightarrow \beta^\prime = \beta + d\lambda
\label{gaugetrans}
\ff
There are six of these per point of $\cal M$, , because $\lambda$ and $\lambda^\prime$ generate the same gauge transform on $\beta$
when $\lambda^\prime = \lambda + df$.  

\subsection{Hamiltonian constrained systems}

To quantize any theory we must first cast it into 
Hamiltonian form\footnote{Some elements of the hamiltonian formulation were
described by Hitchin in \cite{hitchin2}.}. 
Dirac long ago discovered how to construct a hamiltonian system for a theory 
invariant under the diffeomorphisms of a $d+1$ dimensional manifold\cite{carlobook}. 
One considers the manifold to have the form $\Sigma \times R$ where
$\Sigma$ is called the spatial manifold\footnote{There is no need to assume the
degrees of freedom include a  metric, so there
is not necessarily a distinction between timelike and spacelike.}.  
There are $d$ constraints that generate
the diffeomorphisms of $\Sigma$, called ${\cal D}_i$ where $i,j=1,...d$ is a 
spatial index. There is a 
Hamiltonian constraint $\cal H$ that generates the remaining diffeomorphisms in
$\Sigma \times R$.  Any additional gauge symmetries are
generated by constraints ${\cal G}$.  These constraints must form
a first class algebra, which means that they close under Poisson brackets.  

In the spatially compact case, which we will assume here, the Hamiltonian
must be a linear combination of these constraints.  Hence, it must be possible,
by a change of variables, to transform the action to the following form,
\f
I^H= \int dt \int_\Sigma \left (\pi^{ij}\dot{\beta}_{ij} - \rho_a {\cal G}^a
-N^i {\cal D}_i - N {\cal H}  \right )
\label{finalform}
\ff
where $\pi^{ij}$ is the momenta conjugate to $\beta_{ij}$, while $\rho^a$, $N^i$ and $N$ are lagrange multipliers.  

We next proceed to construct such a theory that we conjecture is
equivalent to the theory of Hitchin\cite{hitchin}.

\subsection{Topological $\cal M$ theory as a hamiltonian constrained system}

The action given by Hitchin can be rewritten in a form suggested by 
Eli Hawkins\cite{Eli}.   Let $g_{ab}$ be an arbitrary metric on $\cal M$.  Then
\f
I^{E} [g, \Omega ]= \int_{\cal M} \left [  \sqrt{g} - g^{ab} \tilde{h}_{ab} (\Omega )  \right ]
\label{Eaction}
\ff
gives the same equations of motion as $I^N$ when both $g$ and $\Omega$ are varied. 

Now choose the manifold to be of the form ${\cal M} = \Sigma \times R$
with $\Sigma$ a compact $6$ manifold.     Form indices
in $\Sigma$ will be denoted, $i,j=1,...,6$.  We fix a time coordinate and hence a slicing of $\cal M$ and define canonical momenta
\f
\pi^{ij} = \frac{\delta I^E}{\delta \dot{\beta}_{ij}} 
\label{pidef0}
\ff
where dot denotes as usual derivative by the coordinate on $R$, called $t$.

There are $6$ initial primary constraints given by 
\f
\pi^{0i} = \frac{\delta I^E}{\delta \dot{\beta}_{0i}} =0 
\ff

The Poisson algebra is generated by
\f
\{ \beta_{ij} (x) , \pi^{kl} (y) \} = \delta^{kl}_{ij} \delta^6 (x,y)
\ff
from which we see that the momenta carry density weight one.
This means that the dual is a four-form, $\rho=\pi^*$. In six dimensions, a
four form is  stable (see \cite{hitchin,DGNV} for the meaning of this term) 
and can be written equivalently in terms of a two-form $k$, as 
\f
\rho= k \wedge k 
\ff
The theory may then be expressed equivalently in terms of $k$ or
$\pi^{ij}$. 
For the canonical quantum theory, the latter is more convenient, as
we will see below.

We can see how the action depends on velocities by noting that,
in an obvious notation,
\begin{eqnarray}
\tilde{h}_{ij}& = & (\dot{\beta}_{ij} - d\beta_{oi}) (d\beta_{ij})^2  \nonumber \\
\tilde{h}_{i0}& = & (\dot{\beta}_{ij} - d\beta_{oi})^2  d\beta_{ij}  \nonumber \\
\tilde{h}_{00}& = & (\dot{\beta}_{ij} - d\beta_{oi})^3   
\end{eqnarray}
Hence the action (\ref{Eaction}) is roughly of the form, 
\f
I^E\approx  \int_\Sigma \int dt \left ( (\dot{\beta}_{ij} - d\beta_{oi})^3 A_3
+ (\dot{\beta}_{ij} - d\beta_{oi})^2A_2 + (\dot{\beta}_{ij} - d\beta_{oi}) A_1
+ \mbox{potential}
\right )
\ff
where the $A_I$ are polynomials of spatial derivatives of $\beta_{ij}$.  
As a result we will find an equation of the form
\f
\pi^{ij} \approx (\dot{\beta}_{kl})^2 A_3 +  (\dot{\beta}_{kl}) A_2  + \mbox{constants}
\label{pidef}
\ff
It is not straightforward to invert this relation to find $ \dot{\beta}_{kl}$ as
a function of $\pi^{ij}$.  It may be possible to do this, but for 
the present we proceed by making an educated guess  for the form of the Hamiltonian theory based on our experience with other diffeomorphism invariant systems, and checking its internal consistency as well as its agreement
with known results about Hitchin's action.  
We find such a conjecture, and describe it here.  I believe, but have not shown, that the system of constraints described here, is a restatement of previous
results on this system\cite{hitchin,DGNV}.

We expect
that the inversion of (\ref{pidef}) is only possible modulo a system of
constraints.  This system of constraints will include generators of all
{\it local} gauge invariances of the theory. 

We expect a total of $13$ first class constraints.  Six  will generate
the gauge transformations (\ref{gaugetrans}) in $\Sigma$.  These must have the form,
\f
{\cal G}^i = \partial_k \pi^{ik} =0
\label{gaugec}
\ff
These form an abelian algebra. 

Six constraints will generate local diffeomorphisms of $\Sigma$\footnote{In
\cite{DGNV} and \cite{hitchin} another form of the diffeomorphism constraint
is proposed. It is plausible, but not yet shown, that the two proposed forms
are equivalent, at least on the space of solutions to (\ref{gaugec}). }. 

They will be
given by
\f
{\cal D}_i = \Omega_{ijk} \pi^{jk} =0
\label{diffeo}
\ff
Let us integrate these against a vector field $v^i$, with compact
support on a contractible region of $\Sigma$.
\f
{\cal D}(v) = \int_\Sigma v^i \Omega_{ijk} \pi^{jk}
\ff
It is straightforward to express this as
\f
{\cal D}(v) = \int_\Sigma  \left (
({\cal L}_v \beta_{jk}) \pi^{jk}- 2v^i \beta_{ik} {\cal G}^k 
+ v^i \Omega^0_{ijk} \pi^{jk}
\right )
\ff
If we ignore the last term, then we see that ${\cal D}(v)$
generate a linear combination of diffeomorphisms and
gauge transformations (\ref{gaugetrans}) on $\beta$.  
However, in a compact, topologically trivial region, we can take
$\Omega^0 = d\beta^0$, so that the last term is included in the
previous terms.   It is then straightforward to show that the algebra 
of gauge and diffeomorphism constraints (\ref{gaugec}), and (\ref{diffeo}),
closes, so long as the constraints are multiplied by functions with
support on contractible regions. 

Now we come to the dynamics.  For a diffeomorphism invariant theory on a spatially compact manifold without boundary, this is going to be specified by a hamiltonian constraint $\cal H$, which must be a local density on $\Sigma$.  For such a theory it is a general result that the  Hamiltonian
must be proportional to constraints. The only exception is that there can be
a non-vanishing boundary term, but we are considering here the case of
a manifold without boundary. 

As the action contains terms up to cubic in $\dot{\beta}_{ij}$ we expect
$\cal H$ to have terms up to cubic in $\pi^{ij}$.  
 By analogy with the Ashtekar formalism, we
may expect that the Hamiltonian constraint will be polynomial in the fields
when written as a density of  weight two.   There are two monomials
in $\pi$ and $\Omega$ that give us a scalar of weight two.  
The first is the simplest scalar density 
polynomial in the $\pi^{ij}$, which is,
\f
{\cal K}= \pi^{ij}  \pi^{kl}  \pi^{mn} \epsilon_{ijklmn}
\ff
This is a kind of kinetic energy. For a potential energy we 
seek a scalar of density weight two polynomial in the $\Omega_{ijk}$.
One exists, defined by Hitchin as follows.
Let $\tilde{\kappa}_i^j$ be the densitized, $(1,1)$
tensor,
\f
\tilde{\kappa}_i^j = \Omega_{ikl}\Omega_{mno} \epsilon^{klmnoj}
\ff
Note that the trace $\tilde{\kappa}_i^i =0$. However the trace of the 
square is not zero, and it gives a scalar density of weight two,
\f
{\cal V}= \tilde{\kappa}_i^j  \tilde{\kappa}_j^i 
\ff
Combining them, we have a natural 
candidate  for the hamiltonian constraint\footnote{ This is related to a form of the hamiltonian
studied by Hitchin in \cite{hitchin2}. }, which is
\f
{\cal H}= {\cal K} - a  {\cal V} 
\label{hamiltonian}
\ff
where $a$ is a dimensionless factor. 

We can check this guess by seeing if it leads to a constraint algebra that closes.
The fact that $\cal H$ is  a scalar density of weight two determines that its  Poisson
brackets with (\ref{gaugec}) and (\ref{diffeo}) closes, so long as the gauge
transformations and diffeomorphisms have compact support on contractible
regions. 
To compute the rest of the Poisson algebra we smear against a 
test function $N$ of
density weight minus one, again with compact support in a topologically trivial region. 
\f
{\cal H}(N) =\int_{\Sigma}  N \left (  \pi^{ij}  \pi^{kl}  \pi^{mn} \epsilon_{ijklmn} -
a  \tilde{\kappa}_i^j  \tilde{\kappa}_j^i  \right ) 
\ff
It is straightforward to check that the algebra closes
\f
\{ {\cal H}(N), {\cal H}(M) \} = 
 \int_\Sigma  w^j_{NM} {\cal D}_j = {\cal D}(w_{NM})
\ff
where
\f
w_{NM}^j= 18a (N \partial_i M - M\partial_i N )  \pi^{ik}\tilde{\kappa}_k^j
\ff
Thus, we see that the combination of the $13$ constraints, 
${\cal G}^i$, ${\cal D}_i$ and $\cal H$ make a closed system of first
class constraints.  

In fact, we can argue that its solutions are identical to the solutions
of Hitchin's theory.  When ${\cal H}=0$ we have locally
\f
a \tilde{\kappa}_i^j  \tilde{\kappa}_j^i  =  \pi^{ij}  \pi^{kl}  \pi^{mn} \epsilon_{ijklmn}
\ff
We can take the square root of each side to find that
\f
\sqrt{|a  \tilde{\kappa}_i^j  \tilde{\kappa}_j^i |}   = \sqrt{|\pi^{ij}  \pi^{kl}  \pi^{mn} \epsilon_{ijklmn}|}
\ff
We can find a geometric interpretation of the hamiltonian
constraint.   To do so we note that the $\Omega$ is known to characterize the
complex structure of $\Sigma$ \cite{hitchin,DGNV}.  The densitized bivector
$\pi^{ij}$ provides a symplectic structure.  These fields  allow us to form two different volume elements on the six manifold $\Sigma$.  

There is a volume element associated with the symplectic structure,
\f
\epsilon_\pi =  \sqrt{|\pi^{ij}  \pi^{kl}  \pi^{mn} \epsilon_{ijklmn}|}
\ff
There is similarly a volume element associated with the 
complex structure, given by the three form
metric $h(\Omega)$, pulled back into the six manifold $\Sigma$.  
\f
\epsilon_h =  \sqrt{| \tilde{\kappa}_i^j  \tilde{\kappa}_j^i |} 
\ff
The Hamiltonian constraint says that the two volume forms are
equal to each other, up to the constant $a$.  In \cite{hitchin,DGNV} we see that
Hitchin's  theory implies that 
\f
2\int_{\Sigma} \epsilon_h = \int_\Sigma \epsilon_\pi 
\ff
We see that this condition is implied by the guess for the Hamiltonian
constraint we gave, (\ref{hamiltonian}) so long as $a=\frac{1}{4}$.  
Hence, the diffeomorphism
classes of solutions to the theory given here will coincide with the solutions
of Hitchin's theory.  

Hitchin\cite{hitchin} also provides a translation to the complex geometry
of $6$ manifolds.  He shows (Proposition 2) that when 
\f
{\cal V} <0\
\label{negcondition}
\ff
 one can
define a complex structure
\f
J_i^j = \frac{\tilde{\kappa}_i^j}{\sqrt{-{\cal V}}}
\ff
such that $\Omega$ is the real part of a complex holomorphic three-form. 
  
To summarize we have argued that the Hitchin action can be rewritten as
a constrained hamiltonian system of form, 
(\ref{finalform}) with constraints given by (\ref{gaugec}), (\ref{diffeo}) and
(\ref{hamiltonian}).    We also reach the important conclusion
mentioned in the introduction, that the
complex and symplectic structures are coded by canonically conjugate degrees of
freedom, so long as (\ref{negcondition}) is imposed.  

\section{Counting of degrees of freedom}

It is straightforward to count the local degrees of freedom. 
There are $15$ $\beta_{ij}$ which have $15$ conjugate momenta $\pi^{ij}$.
We have $13$ first class constraints, which will require $13$ gauge fixing
conditions.   
 This leaves $2+2$ canonical degrees of freedom.  Thus the theory is not topological, there are two local degrees of freedom per point of $\Sigma$.  

There are, of course,  also global degrees of freedom, that correspond to integration of
$\Omega$ around non-contractible cycles of $\Sigma$.

\section{Quantization}

Dirac proposed a method to quantize hamiltonian constrained systems. With
some refinements to take into account issues of regularization and ordering that
arise in field theories, this is the method that all background independent
approaches to hamiltonian quantization follow.   
Dirac's method can be further specialized to the case of diffeomorphism
invariant theories whose configuration variables are connections or
$p$-forms with local gauge invariance\cite{carlobook,invitation}.  This specification
of Dirac's method to theories invariant under both diffeomorphisms and local
gauge invariances is the essence of the Hamiltonian part of loop quantum gravity. We first
briefly summarize the procedure, then we apply it to the constrained
system just introduced.

\subsection{Brief review of Dirac quantization}

We begin by specifying  the {\it kinematical 
configuration space}, $\cal C$. In the case of 
topological $\cal M$ theory, this  is the space of two forms $\beta$ on $\Sigma$.
By imposing invariance under the action of  the gauge and spatial diffeomorphism constraints, in this case (\ref{gaugec}) and (\ref{diffeo}), we then  go down to
a gauge and (spatially) diffeomorphism invariant configuration space
\f
{\cal C}^{diffeo} = \frac{\cal C}{\mbox{local gauge transformations} \times \ Diff (\Sigma )}
\ff
The aim of the quantization procedure is to first, construct the corresponding
Hilbert spaces and, second, construct the Hamiltonian constraint as an operator
on diffeomorphism invariant states.

This is accomplished in three steps:

\blankline

{\bf STEP 1}:  Find an algebra $\cal A$ of observables on the kinematical
phase space which has a representation $\hat{\cal A}$ on a Hilbert
space $H^{kinematical}$ such that 

\begin{enumerate}

\item{}The reality conditions of the classical theory, i.e. which variables are real,  are realized by 
the inner product on $H^{kinematical}$.  That is, the inner product is chosen so that
real classical observables are represented by Hermitian operators. 

\item{}$H^{kinematical}$ carries an exact, 
non-anomalous unitary representation
of $Diff (\Sigma)$. This is  given by unitary operators, $\hat{U}(\phi )$, where 
$\phi \in Diff (\Sigma )$.  

\end{enumerate}

\blankline

{\bf STEP 2} Construct a space of diffeomorphism invariant states
$H^{diffeo} \subset H^*_{kinematical}$, which are invariant
under the action of $\hat{U}(\phi )$.  These are the diffeomorphism
invariant states and they live inside the dual of the kinematical Hilbert space.

\blankline

{\bf STEP 3}  Construct a sequence of regularized operators, ${\cal H}^\epsilon (x) $
to represent the Hamiltonian constraints, in $H^{kinematical}$.  Prove that
the limit as $\epsilon \rightarrow 0$ takes diffeomorphism invariant
states to diffeomorphism invariant states, and thus defines 
a finite operator in $H^{diffeo}$.  Prove that the limit
 has a kernel in $H^{diffeo}$ that is infinite dimensional.  
This kernel $H^{physical} \subset H^{diffeo}$ is the physical 
Hilbert space. 

\blankline

When carried out in LQG there are four key observations, that may
extend to the present case

\begin{itemize}

\item{}There is no known way to realize the second condition of 
{\bf STEP 1} when $\cal A$ is the usual local canonical algebra defined
by the gauge connection and conjugate electric fields. In particular,
Fock representations fail because they depend on a background
metric, which breaks diffeomorphism invariance. To proceed one
must base $\cal A$ on extended observables, such as Wilson loops. 

\item{}When $\cal A$ is taken to include the Wilson loops of the
connection, together with conjugate operators linear in
the momenta of the connection,  there is a theorem \cite{LOST} that 
says that there is a unique way to realize the first two steps.  
It is not known if this extends to the present case, but if it does there 
would appear to be only one way to successfully carry out this
program for topological $\cal M$ theory.  

\item{}The kinematical Hilbert space $H^{kinematical}$ is
not separable, because any two distinct, non-overlapping, 
Wilson loop operators 
create orthogonal states. However, this non-separability is exactly
cancelled by imposing diffeomorphism invariance. Hence 
$H^{diffeo}$ is separable, assuming only a technical condition,
which is that it is defined in terms of piecewise smooth diffeomorphisms. 

\item{}When applied to general relativity, all three
steps have been carried out rigorously\cite{rigor}.  

\end{itemize}

\subsection{Quantum topological $\cal M$ theory}

We here sketch how the program just described may be applied
to topological $\cal M$ theory.  We do not attempt to give a rigorous
treatment, but we find that at a particle physics level of rigor we can follow
the same program as was originally used in constructing LQG.  

We begin by finding the algebra of observables analogous to
Wilson loops and their conjugate variables, to represent the local degrees of
freedom.  
We start with the analogue of Wilson loops.   Given any
closed and contractible two surface $S$
in $\Sigma$ we define a function of $\cal C$, 
\f
T[S] = e^{\int_S \beta}
\ff
Similarly, given a four dimensional surface $A \in \Sigma$ we define momentum flux
operators
\f
\Pi [A ] = \int_A \pi^*
\ff
where $\pi^*$ is a four formequivalent to the momenta $\pi$. They have a 
simple Poisson algebra 
\f
\{ T[S], \Pi [A] \} = Int [ S, A] T[S]
\label{algebra}
\ff
where $Int [ S, A]$ is the intersection number of the surfaces $S$ and $A$.

We note that these observables commute with the action of local gauge
transformations generated by (\ref{gaugec}).

Following the strategy just outlined, 
we seek a representation of (\ref{algebra}) on a Hilbert space, ${ H}^{kinematical}$
that carries a nonanamolous representation of $Diff (\Sigma )$.   

Let $\Gamma$ be a network of two surfaces $S \in \Sigma$, whose faces are labeled by integers. The integers count elementary closed surfaces, out
of which the network is formed. This implies that the  triangle inequalities are satisfied
at every trivalent edge where surfaces meet.  

 States are functionals of $\Gamma$, so we have
\f
<\Gamma |\Psi > = \Psi (\Gamma )
\ff
The operator representing $T[S]$ is defined by
\f
<\Gamma | \circ \hat{T}[S]= <\Gamma \oplus S |
\ff
where $\Gamma \oplus S $ is the network $\Gamma$ with the surface $S$ added.  
This gives us
\f
 \hat{T}[S] \circ \Psi [\Gamma ] = \Psi [\Gamma \oplus S ]
\ff
The conjugate momentum operator $\Pi [A]$ is defined by
\f
<\Gamma | \hat{\Pi}[A] = \imath \hbar \sum_{S\in \Gamma } Int[S,A] <\Gamma |
\ff
One can check explicitly that the commutator
\f
[ \hat{T}[S], \hat{\Pi} [A] ] =\imath \hbar Int [ S, A] \hat{T}[S]. 
\ff
The kinematical inner product is
\f
<\Gamma |\Gamma^\prime > = \delta_{\Gamma \Gamma^\prime}
\label{kinIP}
\ff
This gives rise to a non-separable Hilbert space, as in LQG. 
The unitary representation of $\phi \in Diff (\Sigma )$ is defined by
\f
<\Gamma | U(\phi ) = <\phi \circ \Gamma |
\ff
This is easily shown to be unitary under (\ref{kinIP}). 

We then define the diffeomorphism invariant Hilbert space, ${ H}^{diffeo}$ to be those 
states in the dual of ${ H}^{kin}$ such that
\f
<\Psi | \hat{U}(\phi ) = <\Psi |
\ff
Following the standard method of LQG, these can be shown to have a countable, orthonormal basis, given by $|\{ \Gamma  \}  >$, where $\{ \Gamma  \}$ are diffeomorphism classes\footnote{To eliminate continuous labels on states coming from
labeling diffeomorphism  equivalence classes of complicated intersections, 
the diffeomorphisms are extended to piecewise
smooth diffeomorphisms, after which the basis is countable\cite{carlobook}.  } of networks of labeled two-surfaces embedded in
$\Sigma$

We now want to introduce a regularized Hamiltonian constraint operator,
$\hat{\cal H}_\epsilon$
expressed in terms of elements of the surface algebra, in the kinematical
Hilbert space.  This should have several properties:

\begin{enumerate}

\item{} On the classical counterpart, 
$\lim_{\epsilon \rightarrow 0 }{\cal H}_\epsilon = {\cal H}$.

\item{} The limit $\lim_{\epsilon \rightarrow 0 }\hat{\cal H}_\epsilon$
acts on $H^{diffeo}$ in that it takes diffeomorphism invariant states to
diffeomorphism invariant states.

\end{enumerate}

Here are some steps towards the construction of such a regularized operator.
A regularization procedure is going to break diffeomorphism invariance in 
$\Sigma$.  
So let us introduce in a local region $\cal R$, a flat metric $q^0_{ij}$ in $\Sigma$ and a set of coordinates $y^i$.  At a point $p \in \Sigma$ we
can have a box ${\cal B}_{\hat{i}\hat{j}\hat{k}}^{\epsilon} (p)$ of Volume
$\epsilon^3$ in $q^0_{ij}$ alongside the $\hat{i}, \hat{j}\hat{k}$ axis.
We have, to leading order
\f
T[{\cal B}_{\hat{i}\hat{j}\hat{k}}^{\epsilon} (p)] = 1 + \epsilon^3 F_{\hat{i}\hat{j}\hat{k}} (p) 
\ff
where $T[{\cal B}_{\hat{i}\hat{j}\hat{k}}^{\epsilon} (p)]$ takes the intergral of 
$\beta$ around the surface of the box.  

We can then write a regularized three form operator as
\f
\Omega^\epsilon_{\hat{i}\hat{j}\hat{k}}= \frac{1}{\epsilon^3} 
\left (
T[{\cal B}_{\hat{i}\hat{j}\hat{k}}^{\epsilon} (p)] -1
\right )
\ff
We can then write a regulated Hamiltonian constraint operator
\f
{\cal H}^\epsilon = {\cal K}^\epsilon + {\cal V}^\epsilon
\ff
with
\f
 {\cal V}^\epsilon= \tilde{\kappa}_i^{\epsilon j}  \tilde{\kappa}_j^{i \epsilon} 
\ff
where the regulated operator $ \tilde{\kappa}_j^{i \epsilon} $ is 
\f
 \tilde{\kappa}_{\hat{i}}^{\hat{j} \epsilon} = \Omega^\epsilon_{\hat{i}\hat{k}\hat{l}}
\Omega^\epsilon_{\hat{m}\hat{n}\hat{o}}  
\epsilon^{\hat{k}\hat{l} \hat{m}\hat{n}\hat{o}\hat{j} } 
\ff
Similarly, let us define a surface $A^{\hat{i}\hat{j}}_\epsilon (p)$  to be a four dimensional hypercube of size $\epsilon$ on a side, 
orthogonal to the $\hat{i}\hat{j}$  directions, all with respect to the background
metric $q^0_{ij}$, at the point $p$. We then can define
\f
\hat{\Pi}^{\hat{i}\hat{j}}_\epsilon= \frac{1}{\epsilon^4} \Pi (A^{\hat{i}\hat{j}}_\epsilon (p) ) 
\ff
We then  have for the regularized kinetic energy 
\f
{\cal K}^\epsilon =  
\epsilon_{\hat{i}\hat{j}\hat{k}\hat{l}\hat{m}\hat{n}}
\hat{\Pi}^{\hat{i}\hat{j}}_\epsilon   \hat{\Pi}^{\hat{k}\hat{l}}_\epsilon \hat{\Pi}^{\hat{m}\hat{n}}_\epsilon  
\ff

There remains much to do, but the outline is clear from here, by analogy with the development of LQG. For example, one can define a path integral by
exponentiation.  It will be defined as a spin foam model, based on labeled
triangulations of $\cal M$. The three-simplices of the triangulation will be
labeled with integers, corresponding to the evolutions of the graphs.  There
will also be labels on the four-simplices, corresponding to $\pi^*$.

\section{Down from $11$ dimensions}

We do not have a background independent formulation of $\cal M$ theory, so the existence of the theory remains a conjecture.  But part of that conjecture is 
that a classical limit of $\cal M$ theory is given  by $11$ dimensional supergravity.  Hence it is of interest to see if  Hitchin's $7$ dimensional theory might be derived from a suitable reduction of $11$ dimensional supergravity.  If it can be, then it may be that
we can identify the quantum states just described as the actual quantum
degrees of freedom corresponding to the membranes of $\cal M$ theory. 

As is the case in all known versions of general relativity and supergravity, the action and field equations 
for $11$ dimensional
supergravity can be written in a polynomial form \cite{d=11}.   
This makes it possible to take a consistent
reduction in which the frame field, connection and gravitino fields (with certain density weights) 
are taken to zero, leaving only the
three form $a_{ABC}$\footnote{$A,B,...= 0, 1,...,10$, while ten dimensional spatial indices
are given by $I,J,...= 1,...,10$.}   Since the field equations are polynomial, these provide a subset of
solutions to the full equations of $11d$ supergravity.  It can be shown that the supersymmetry transformations, which are also polynomial, are trivially satisfied for such solutions.  

In this reduction, the action is 
\f
I^{11} = \int_{{\cal M}^{11}} da \wedge da\wedge a
\label{11d}
\ff
This is a version of higher dimensional Chern-Simons theory. Its dynamics and quantization were studied in detail
in \cite{d=11}.  
It is important to note that higher dimension and higher form Chern-Simons theories have
local degrees of freedom. This theory is diffeomorphism invariant, but it is not topological.

Let us see if  the degrees of freedom of  Hitchin's $7$ dimensional theory can be
found imbedded in this metric-less reduction of $11$ dimensional supergravity.  

The field equations of (\ref{11d}) are
\f
da \wedge da =0
\label{11eq}
\ff
The gauge transformation is of course
\f
\delta a = d\lambda
\label{11gauge}
\ff
with $\lambda$ a two form.

A solution to (\ref{11eq}) is given by the following ansatz. Let $A=a,\alpha$, with $a=0,...,d$
and $\alpha = d+1,...10$, Then the ansatz is
\f
da_{\alpha ABC } =0
\ff
so long as $d \leq 6$.  It is interesting that the largest non-trivial case is $d=6$, which gives us a 
reduction to a seven dimensional theory.

In fact, we can find a simple set of solutions, that are locally but not globally flat.
Let the topology be chosen to be the standard one proposed for a reduction from
$\cal M$ theory to string theory, 
\f
{\cal M}^{11} = R \times \Sigma \times S^1 \times R^3 . 
\ff
Here $\Sigma$ is a compact six manifold, and the $R$ is time, as in previous sections.    
These are coordinatized as before by $x^a$, $a=0,i$, with $i=1,...,6$.  
Let $y^7=\theta$ be
the coordinate around the $S^1$.  This is the standard circle around which membranes
are wrapped to get strings.  The three remaining dimensions in the $R^3$ can as usual be taken to 
be ordinary, uncompactified space, coordinatized by $y^\alpha$ with $\alpha=1,2,3$ from now on.  
We will  assume that everything is constant in space, so that
\f
\frac{\partial a_{ABC}}{\partial y^\alpha} =0
\ff
This, physically, means that we are studying the geometry of string compactifications 
that might arise from $\cal M$ theory.

Let us take a solution which is locally pure gauge, of the form of (\ref{11gauge}), with
(locally on the $S^1$)
\f
\lambda_{ab}= \theta \gamma_{ab} (x) 
\ff
However globally, we will have
\f
\beta_{ab} (x) = \oint_{S^1} d\theta a_{\theta ab} = \gamma_{ab} (x) 
\ff
Since the solution is locally trivial, $da=0$ everywhere, so this is a solution to $11d$ supergravity.

From (\ref{11gauge}) we see that there  is still a gauge invariance, given by 
\f
\delta \beta = \delta \gamma = d\phi
\ff
where $\phi$ is a one form. 

The integral around three-cycles $C_I$ of $\Sigma$ are given by 
\f
\int_{C_I} a = \int_{C_I} d\beta  
\ff
These are constants, as they do not evolve in time under the equations of motion (\ref{11eq}). 

The canonical momenta for $a_{IJK}$ is $\Pi^{IJK} \sim ( a\wedge da )^* $, where
the duality is in the ten dimensional manifold $\Sigma \times S^1 \times R^3$.
We have
\f
\{ a_{IJK} ( x ), \Pi^{LMN}(y) \} = \delta^{10} (x,y) \delta_{IJK}^{LMN}
\ff
The dimensionally reduced momenta is
\f
\pi^{*} = \int_{R^3 } \Pi^*
\ff
which is a four-form on $\Sigma \times R$. It can be pulled back to a
four form on $\Sigma$.   We have
\f
\{  \beta_{ij}   (x^i ) ,  \pi^{*}_{klmn} (y^i )  \} = \int_{S^1} d\theta 
\int_{R^3} d^3x^{\alpha \beta \gamma }
\{ a_{\theta ij} , \Pi^*_{klmn \alpha \beta \gamma } \} = \delta^6 (x^i, y^i )
\epsilon_{ijklmn}
\ff
Thus, the canonical degrees of freedom of topological $\cal M$ theory can
be seen to arise from the reduction of supergravity from $11$ dimensions. 

The reduced theory then has degrees of freedom $(\beta_{ab}, \pi^{cd})$, with fixed cohomology on
$R\times \Sigma$.  In the quantum theory there will arise an effective action to describe
the low energy dynamics of these degrees of freedom.  The effective action will be dominated by the lowest dimension term that can be made from $d \beta$
on $\Sigma \times R$.  One can 
conjecture that this will  be given by the Hitchin's action, which
is a cosmological constant term, and hence should dominate the low energy limit.

It is possible we can proceed further in this direction. 
Let $g$ be a flat metric on the $R^4$ parameterized by
$x^0$ and $y^\alpha$ and let $e^0, e^\alpha$ be four one form orthonormal frame fields.  
Given the imbedding of $R^4$ into ${\cal M}^{11}$ we can pull these back to a degenerate
set of $11$ dimensional frame fields.  We may conjecture that these, together with any $a_{ABC}$
such that locally $da=0$, give solutions of $11$ dimensional supergravity. If this is true, then    there is a sector of $\cal M$ theory 
with a conventional geometry on the four uncompactified spacetime dimensions, but where the geometry on the
compactified dimensions is entirely based on a forms theory.

There should be much more in this sector.  We should be able to add other degrees of freedom coming from the fields of $11$ dimensional supergravity to systematically expand Hitchin's theory to a reduction of
$\cal M$ theory, with a full set of local degrees of freedom.  

\section{Conclusions}

What is described here is a first step towards a background independent quantization
of topological $\cal M$ theory.  Many issues remain open.  While the conjecture that the
constrained system here is equivalent to Hitchin's seven dimensional theory is plausible,
it still needs to be proved.  The results on the quantum theory are just a first sketch,
along the lines of early papers on LQG.  It is likely that the quantization can
be made rigorous, along the lines of \cite{rigor}.  Of great interest is whether
there is an extension of the LOST uniqueness theorem\cite{LOST} to this
context.  Further exploration of this direction can be expected to shed light both on the key question of what $\cal M$ theory may be as well as on the interpretation of the results
in $LQG$ concerning $3+1$ dimensional physics.  The idea proposed in \cite{DGNV}
that Hitchin's theory may open the way to a unification of string theory and $LQG$ is
intriguing and the results obtained here give us a common language within which the
precise relationship between the two approaches can be elucidated. 

But we can already draw a few interesting conclusions from the results obtained here. 
First we see a possible non-perturbative origin for D-brane states in a background
independent formulation.  Second, as pointed out in \cite{DGNV}, there are implications for string 
compactifications.  
In the standard string compactifications on Calabi-Yau manifolds, the $\Omega$ and $\pi^{ij}$ are fixed. 
However, we see that in topological $\cal M$ theory these are conjugate variables. Moreover, we see that these variables can be understood to descend from full $\cal M$ theory, where they are still conjugate variables. If so, then  there can be no quantum states of $\cal M$ theory corresponding to fixed
Calabi Yau geometries on $\Sigma$.  Thus, any phenomenology that depends on the fixed background structure of a Calabi-Yau manifold can only be meaningful in the semiclassical limit in which conjugate variables can both have definite values.  

Finally, it is interesting to note that the real variables on which Hitchin's theory is based only define
a complex manifold when the condition (\ref{negcondition}) is satisfied. 
This is analogous to the condition that the determinant of the spatial metric
be positive.  It means that the part of the configuration space that corresponds
to complex geometries is not a vector space, but
satisfies a non-linear inequality.  There is then the issue of how this
inequality is to  be satisfied in the quantum theory.   Just as the metric may have an
amplitude to be 
non-degenerate in any first order formulation of quantum gravity, so we must consider
the possibility that a quantum state can give a non-zero amplitude to a region of
configuration space where (\ref{negcondition}) is violated, leading to quantum
fluctuations in which $\Sigma$ fails to have a complex structure.

\section*{ACKNOWLEDGEMENTS}

I am grateful to Robert Dijkgraaf, Sergei Gukov,  Nigel Hitchin, Andrew Neitzke 
and Cumrun Vafa for taking the time to explain to me the basics
of topological string and $\cal M$ theory, and especially to Andrew Neitzke for
comments on the manuscript.   I am also very grateful to Christian Romelsberger and Jaume Gomis for very helpful comments and, especially, to  Eli Hawkins for  many perceptive and useful suggestions. Finally, thanks to Stephon Alexander for suggestions which improved the manuscript.

\end{document}